\documentclass{preprint}
\usepackage{amsmath}
\usepackage{amsfonts}
\usepackage{amsbsy}
\usepackage{times}
\usepackage{mhequ}
\usepackage{mhenvs}
\usepackage{ScriptFonts}
\newcommand{\LL}{{\mathcal L}}

\newcommand{\KK}{{\mathcal K}}
\newcommand{\F}{{\mathcal F}}
\newcommand{\T}{{\mathbb T}}
\newcommand{\R}{{\mathbb R  }}

\def\eps{\varepsilon}
\let\epsilon\varepsilon
\def\eref#1{(\ref{#1})}

\def\n#1{\left|\!\left|\!\left|#1\right|\!\right|\!\right|}

\def\x(#1){x^{\eps}_{#1}}
\def\y(#1){y^{\eps}_{#1}}
\def\xe(#1){\eps x^{\eps}_{#1}}
\def\cS{{\cal S}}
\def\CP{{\cal P}}
\def\CC{{\cal C}}
\def\CH{{\cal H}}

\newcommand\scal[2][ ]{\ifthenelse{\equal{#1}{ }}{\langle#2\rangle}{}
        \ifthenelse{\equal{#1}{b}}{\bigl\langle#2\bigr\rangle}{}
        \ifthenelse{\equal{#1}{B}}{\Bigl\langle#2\Bigr\rangle}{}
        \ifthenelse{\equal{#1}{bb}}{\biggl\langle#2\biggr\rangle}{}
        \ifthenelse{\equal{#1}{BB}}{\Biggl\langle#2\Biggr\rangle}{}}
\def\expect{\mathbf{E}}

\def\eg{\textit{e.g.}}
\def\prob{\mathbf{P}}
\def\TV{{\mathrm{TV}}}
\def\d{\partial}
\def\L^#1{\mathrm{L}^{\!#1}}

\newcommand{\BB}{{\cscr B}}
\newcommand{\C}{{\cscr C}}
\begin{document}
\title{Periodic Homogenization for Hypoelliptic Diffusions}
\author{M. Hairer$^{\hbox{\rm{\scriptsize 1}}}$ and G. A. Pavliotis$^{\hbox{\scriptsize\rm 2}}$}
\institute{Mathematics Institute, Warwick University\\
        Coventry, CV4 7AL, United Kingdom\\
        $^{\hbox{\rm{\tiny 1}}}$Email: \texttt{hairer@maths.warwick.ac.uk}\\
        $^{\hbox{\rm{\tiny 2}}}$Email: \texttt{pavl@maths.warwick.ac.uk}
                    }
\maketitle
\begin{abstract}
We study the long time behavior of an Ornstein--Uhlenbeck process
under the influence of a periodic drift. We prove that, under the standard diffusive
rescaling, the law of the particle position converges weakly to the law of a Brownian
motion whose covariance can be
expressed in terms of the solution of a Poisson equation. We also derive upper bounds on the
convergence rate.
\end{abstract}

\section{Introduction}
\label{sec:intro}
In this paper we study the long time behavior of solutions of the following Langevin
equation:

\begin{equation}
\tau \ddot{x}(t) = v(x(t)) - \dot{x}(t) + \sigma \dot{\beta}(t)\;,\quad x(t) \in
\R^n\;,
\label{eqn:intro}
\end{equation}
where $\beta(t)$ is a standard Brownian motion and $\sigma, \, \tau >0$. The parameter
$\tau$ can be thought of as a nondimensional particle relaxation time, which measures
the inertia of the particle. The drift term
$v$ is taken to be smooth, periodic with period $1$ in all directions; further, it is
assumed that it satisfies an appropriate centering condition.

It is well known that as $\tau$ tends to $0$ the solution of \eqref{eqn:intro}
converges with probability $1$ to the solution of the Smoluchowski equation
\begin{equation}
\dot{z}(t) = v(z(t)) + \sigma \dot{\beta}(t)\;,\quad x(t) \in
\R^n\;,
\label{eqn:smol}
\end{equation}
uniformly over every finite time interval, see e.g. \cite[Ch~10]{nelson}.
The problem of homogenization for equation \eqref{eqn:smol} has been studied
extensively over the last three decades for periodic \cite{lions, bhatta, pardoux} as
well as random \cite{carmona, komor_olla, landim_olla} drifts. For the case where
$v(z)$ is a smooth, periodic field which is centered with respect to the invariant
measure of the process, it is not hard to prove \cite[Ch~3]{lions} that the rescaled
process $\epsilon z(t/\epsilon^2)$ converges, as $\epsilon$ tends to $0$, to a
Brownian motion with a positive definite covariance matrix $\KK$. The proof of this
functional central limit theorem is based on estimates on the spectral gap of the
generator of the process $z(t)$.

The long time behavior of particles with non--negligible inertia, whose evolution is
governed by equation \eqref{eqn:intro} has been investigated by Freidlin and coworkers in
a series of papers \cite{freidlin1, freidlin4, freidlin5, freidlin7}. Among other things,
 Hamiltonian systems under weak deterministic and random perturbations were studied in
 these papers:
\begin{equation}
\tau \ddot{x} = - \nabla V(x) + \epsilon (- \kappa \dot{x} + \gamma) +
\sqrt{\epsilon} \sigma \dot{\beta},
\label{e:freid}
\end{equation}
with $\kappa, \, \gamma \in \R$. It was shown that, under appropriate assumptions on
the potential $V(x)$, the rescaled process $\{x(t/ \epsilon), \, y(t / \epsilon) \}$
converges weakly to a diffusion process on a graph corresponding to the Hamiltonian
of the system $H = \frac{1}{2} \tau \dot{x}^2 + V(x)$.

On the other hand, the problem of homogenization for \eqref{eqn:intro} has been
investigated less. This is not surprising since the hypoellipticity of the generator
of the process \eqref{eqn:intro} renders the derivation of the necessary spectral
gap estimates more difficult. Homogenization results for the solution $x(t)$ of
\eqref{eqn:intro} have been obtained, to our knowledge,  only for the case where the
drift $v(x)$ is the gradient of a potential. In this case the invariant measure of
the process $\{x(t), \, \dot{x}(t) \}$ is explicitly known and this fact simplifies
considerably the analysis. This problem was analyzed for periodic \cite{rodenh} as
well as random potentials \cite{papan_varadhan}. In both cases it was shown that the
particle position converges, under the diffusive rescaling, to a Brownian motion
with a positive covariance matrix $\KK$. The proofs of these homogenization theorems
are based on the techniques developed for the study of central limit theorems for
additive functionals of Markov processes \cite{kipnis}, together with a
regularization procedure for appopriate degenerate Poisson equations. Related
questions for subelliptic diffusions have also been investigated
\cite{norris1,norris2, batty}.

The purpose of this paper is to prove a central limit theorem for the solution of
the Langevin equation \eqref{eqn:intro} with a general periodic smooth drift $v(x)$
and, further, to obtain bounds on the convergence rate. The proof of our
homogenization theorem relies on the strong ergodic properties of hypoelliptic
diffusions. The techniques developed in \cite{EPR,EH00} enable us to prove the
existence of a unique, smooth invariant measure for \eqref{eqn:intro} and to obtain
precise estimates on the solution of the Poisson equation $-\LL f = g$, where $\LL$
is the generator of the process \eqref{eqn:intro} and the function $g$ is smooth and
centered with respect to the invariant measure. Based on these estimates it is
rather straightforward to show that the rescaled particle position $\epsilon
x(t/\epsilon^2)$ convergences to a Brownian motion, using the techniques developed
in \cite{kipnis}. Obtaining bounds on the rate of convergence requires more work.
For this, we need to identify the limiting Brownian motion and to introduce an
additional Poisson equation.

The sequel of this paper is organized as follows. In section \ref{sec:thm} we introduce the
notation that we will be using throughout the paper and we present our main result,
Theorem~\ref{thm:conv}. In section \ref{sec:est} we prove various estimates
on the invariant measure of \eqref{eqn:intro} and the solution of the cell problem,
and we also derive estimates on moments of the particle velocity. The proof of
Theorem \ref{thm:conv} is presented in section \ref{sec:proof}. Finally, section
\ref{sec:concl} is reserved for a few concluding remarks.

\begin{acknowledge}
We would like to thank Sigurd Assing, Xue-Mei Li, Neil O'Connell, Andrew Stuart, and Roger Tribe
for useful discussions and comments. We are also grateful to the MRC at the University
of Warwick for
their warm hospitality. The research of MH is supported by the Fonds National Suisse.
\end{acknowledge}

\section{Notation and Results}
\label{sec:thm}
Consider the following Langevin equation in $\R^n$:
\begin{equation}
\tau \ddot{x}(t) = v(x(t)) - \dot{x}(t) + \sigma \dot{\beta}(t)\;,
\label{eqn:langevin}
\end{equation}
with initial conditions $x(0) = x, \, \dot{x}(0) = ( \sqrt\tau )^{-1} y$. We assume
throughout this paper that $v \in \CC^{\infty}(\T^n)$. Introducing $y(t) = \sqrt\tau \dot x(t)$, we
rewrite \eref{eqn:langevin} as a first order stochastic differential equation:
\begin{equa}[e:system]
dx(t) &= {1\over \sqrt\tau} y(t)\,dt \, ,  \\
dy(t) &= {1\over \sqrt\tau} v(x(t))\,dt - {1\over \tau}y(t)\,dt +{\sigma \over
\sqrt\tau}d\beta(t)\;.
\end{equa}
We denote  by  $\LL$  the generator of the process $\{x(t), y(t) \}$:
\begin{equation}
\mathcal{L} = \frac{1}{\sqrt{\tau}} \left( y \cdot \nabla_x + v(x) \cdot \nabla_y
\right) + \frac{1}{\tau} \left(- y \cdot \nabla_y + \frac{\sigma^2}{2} \Delta_y
\right).
\label{eqn:generator}
\end{equation}
By Theorem \ref{theo:boundIM} below, the process $\{x(t), y(t) \}$ admits a unique,
smooth invariant measure, denoted by $\mu(dx,dy)$.

Consider now the {\it cell problem}
\begin{equation}
-\LL \Phi = \frac{1}{\sqrt{\tau}} y\;.
\label{eqn:cell}
\end{equation}
This equation has a unique, smooth solution in the appropriate function space by
Theorem \ref{thm:boundphi} , provided that $\int v(x) \, \mu(dx, dy) = 0$. We define
the symmetric, positive $n \times n$ matrix $\KK$ such that
\begin{equation}
\KK^2 = \frac{\sigma^2}{\tau} \int \nabla_y \Phi \otimes \nabla_y\Phi \,d\mu\;.
\label{eqn:cov}
\end{equation}
The main result of this paper is that the particle position, under the standard
diffusive rescaling, converges weakly to a Brownian motion with covariance $\KK^2$.
We furthermore give upper bounds on the rate of convergence in the following metric.
Let $\BB$ denote a separable Banach space and $\BB^*$ be its dual space. Given two
measures $\mu_1$ and $\mu_2$ on $\BB$, we also denote by $\C(\mu_1, \mu_2)$ the
set of all measures on $\BB^2$ with marginals $\mu_1$ and $\mu_2$. With these notations,
we define the following metric on the space of probability measures on $\BB$ with finite
$p$-moment:
\begin{equ}[e:weakW]
\n{\mu_1 - \mu_2}_p^p = \sup_{\ell \in \BB^*} \inf_{\mu_\ell \in \C(\mu_1, \mu_2)}
\int_{\BB^2} {|\ell(x) - \ell(y)|^p\over \|\ell\|^p}\,\mu_\ell(dx, dy)\;.
\end{equ}
This distance is close in spirit to the $p$-Wasserstein distance
\begin{equ}
\n{\mu_1 - \mu_2}_{p,W}^p = \inf_{\mu \in \C(\mu_1, \mu_2)}
\int_{\BB^2} \|x - y\|^p\,\mu(dx, dy)\;,
\end{equ}
so we will refer to it as the weak $p$-Wasserstein distance. Note that the distance \eref{e:weakW} gives
a locally uniform bound on the distance between characteristic functions
$\chi_\mu(\ell) = \int e^{i\ell(x)}\,\mu(dx)$:
\begin{equ}
|\chi_{\mu_1}(\ell) - \chi_{\mu_2}(\ell)| \le \|\ell\|\,\n{\mu_1 - \mu_2}_p\;.
\end{equ}
In particular one has $\n{\mu_1 - \mu_2}_p = 0$ if and only if $\mu_1 = \mu_2$.

In order to simplify notations, we define the fast processes $\y(t) = y(\eps^{-2}t)$
and $\x(t) = x(\eps^{-2} t)$. We will also from now on use the notation $\BB = \CC([0,T], \R^n)$, for
a value $T > 0$ that remains fixed throughout this paper.
Now we are ready to state the homogenization theorem.
%
% Main theorem
%
\begin{theorem}
\label{thm:conv}
Let $x(t)$ be the solution of \eqref{eqn:langevin}, in which the velocity field
$v \in \CC^{\infty}(\T^n)$ satisfies $\int v(x) \, \mu(dx, dy) = 0$. For $T > 0$ fixed, denote by
$\mu_\eps$ the measure on $\BB$ given by the law of the rescaled process $\xe(t)$
and by $\mu$ the law of a Brownian motion on $\R^n$ with covariance $\KK^2$ as defined in
\eref{eqn:cov}.
Then, for  every $p \ge 1$ and $\alpha \in \left(0, \frac{1}{2} \right)$, there is
a constant $C$ such that
\begin{equ}[eqn:est]
\n{\mu_\eps - \mu}_p \leq C \eps^{\alpha}\;,
\end{equ}
for all $\eps \in (0,1)$. Furthermore, if one denotes by $\pi_k:\BB\to\CC([0,T], \R)$ the
projection given by $\bigl(\pi_k x\bigr)(t) = \scal{k, x(t)}$, one has the bound
\begin{equ}[e:strong]
\n{\pi_k^*\mu_\eps - \pi_k^*\mu}_{p,W} \leq C \eps^{\alpha}\;,
\end{equ}
for every $k\in \R^n$ with $\|k\|\le 1$.
\end{theorem}

\begin{remark}
The condition $\int v(x) \, \mu(dx, dy) = 0$ ensures that there is no ballistic motion involved.
In the general case, one can write $\bar v = \int v(x) \, \mu(dx, dy)$ and define
$\xe(t) = \eps x(\eps^{-2} t) - \eps^{-1} \bar v t$. Then, Theorem \ref{thm:conv} holds for
$\xe(t)$.
\end{remark}

\begin{remark}
If $n = 1$, the bound \eref{e:strong} is much stronger than the bound \eref{eqn:est}. If $n > 1$
however, this bound does not imply any form of convergence $\mu_\eps \Rightarrow \mu$.
It is indeed possible to construct two Gaussian stochastic processes $x(t)$ and $y(t)$ with values in $\R^2$ such
that the law of $x$ differs from the law of $y$ and such that, for every $k \in \R^2$, the law
of $\scal{k,x}$ is identical to the law of $\scal{k,y}$. As an example, choose three i.i.d.\ Gaussian
centered random variables $a_1, a_2, a_3$ and define
\begin{equs}[4]
x_1(t_1) &= a_1 \quad&\quad x_2(t_1) &= a_2 \quad&\quad
x_1(t_2) &= a_3 \quad&\quad x_2(t_2) &= a_1\\
y_1(t_1) &= a_1 \quad&\quad y_2(t_1) &= a_2 \quad&\quad
y_1(t_2) &= a_2 \quad&\quad y_2(t_2) &= a_3 \;.
\end{equs}
It is an easy exercise to check that  these two processes possess the required properties.
\end{remark}

\begin{remark}
Convergence in the weak $p$-Wasserstein distance alone does \textit{not} imply weak convergence,
as the space of probability measures on $\BB$ is not complete under $\n{\,\cdot\,}_p$. This can be seen
by taking $\BB = \ell^2$ and choosing for $\mu_n$ the Gaussian measure with covariance
\begin{equ}
Q_n = \mbox{diag} \bigl(1, {\textstyle{1\over 2}}, \ldots, {\textstyle{1\over n}},0,\ldots\bigr)\;.
\end{equ}
It is straightforward to check that this forms a Cauchy sequence with respect to $\n{\,\cdot\,}_p$, but
does not converge to any measure supported in $\ell^2$. (It does however converge weakly to
a limiting measure in a weaker topology, and this is always the case.) In our case, it is easy to check that
the sequence of measures $\mu_\eps$ is tight, since the generalized Kolmogorov criteria
\cite[Thm~2.1]{yor} provides us with uniform bounds on the $\alpha$-H\"older constant (with $\alpha < {1\over 2}$) of the process $x_\eps$.
Tightness, together with convergence in the weak $p$-Wasserstein distance then implies weak convergence. Note also that even though convergence in the weak $p$-Wasserstein distance
alone does not imply weak convergence, it does imply
weak convergence of finite-dimensional distributions.
\end{remark}

\begin{remark}
The covariance, or {\it effective diffusivity}, $\KK^2$ of the limiting Brownian motion depends on
the $\sigma$ and $\tau$. It is shown in \cite{per_hom_inert_part} that as $\tau$ tends to $0$ the
covariance $\KK^2$ converges to the one obtained from the homogenization of equation \ref{eqn:smol}.
 We refer to \cite{per_hom_inert_part} for further properties of the effective
 diffusivity, together with numerical experiments for various fields $v(x)$.
\end{remark}

\begin{remark}
For simplicity, we choose the molecular diffusion $\sigma$ to be a constant scalar. Taking for
$\sigma$ a positive definite matrix would only require a slight change in our notations.
We could even allow $\sigma$ to depend on $x$ in a smooth way, as long as it remains
strictly positive definite for all $x \in \T^n$. The results
from \cite{EPR,EH00} then still apply and one can check that all the bounds obtained in section~\ref{sec:est} still hold. Since the proof of \theo{thm:conv} itself never uses the fact that $\sigma$ is constant,
all of our result immediately carry over to this case.
\end{remark}

\begin{remark}
For simplicity, we assumed the initial condition $(x,y)$ to be deterministic. However, it is
easy to check that all our arguments work for randomly distributed initial conditions provided
that they are independent of the driving noise and that $\expect \exp \delta \|y\|^2 < \infty$ for
all $\delta \in (0, \sigma^{-2})$. In particular, one can take the initial condition to be distributed
according to the invariant measure.
\end{remark}
The proof of this theorem will be presented in section \ref{sec:proof}.

%The first step of the proof
%is to apply It\^{o} formula to the solution $\Phi$ of the cell problem \eref{eqn:cell}. This enables
%us to write the process $\xe(t)$ as the sum of a ``small'' term, plus a martingale part
%$M^{\epsilon}_t$.  This a quite standard
%method for studying additive functionals of Markov processes, see e.g. \cite{kipnis}.
%Next, we use the Dambis--Dubins--Schwartz theorem \cite[Thm~1.6]{yor} to write
%the martingale $M^{\epsilon}(t)$ as a time change of a Brownian motion $B(t)$.
%In order to do
%so, we need to introduce an additional cell problem, equation \eqref{eqn:F} below. We
%also need the estimates on the moments of the particle velocity presented in section
%\ref{sec:vel}.
%
\section{Preliminary Estimates}
\label{sec:est}
In this section we collect various estimates which are necessary for the proof of the homogenization
theorem. In section \ref{subsec:phi} we study the structure of the invariant measure $\mu$ for
\eqref{eqn:langevin}. We show that it possesses a smooth density with respect to the
Lebesgue measure and we derive sharp bounds for it. Further, we investigate the
solvability of the Poisson equation
\begin{equation}
- \LL f = h,
\label{e:pois}
\end{equation}
where $h$ is a smooth function of $x$ and $y$ which is centered with respect to $\mu$. We prove that
equation \eqref{e:pois} has a smooth solution which is unique in the class of functions which do not
grow too fast at infinity.

In section \ref{sec:vel} we derive estimates on exponential moments of the particle velocity.
Roughly speaking, these estimates imply that the particle velocity grows very slowly with time.
\subsection{Bounds on the invariant measure and on the solution of the Poisson equation}
\label{subsec:phi}
If $v = 0$, the invariant measure for \eref{eqn:langevin} is given by $\mu =
e^{-{\|y\|^2 \over \sigma^2}}\,dx\,dy$. This is ``almost'' true also in the case $v
\neq 0$, as can be seen by the following result.

\begin{theorem}\label{theo:boundIM}
Let $\mu$ be the invariant measure for \eref{eqn:langevin} and denote by $\rho(x,y)$
its density with respect to the Lebesgue measure. Then, for every $\delta \in (0,2\sigma^{-2})$ one can write
\begin{equ}[e:defrho]
\rho(x,y) = e^{-{\delta\over 2} \|y\|^2} g(x,y)\;,\qquad g\in\cS\;,
\end{equ}
where $\cS$ denotes the Schwartz space of smooth functions with fast decay.
\end{theorem}

\begin{proof}
The proof follows the lines of \cite{EPR,EH00}. Denote by $\phi_t$ the (random) flow
generated by the solutions to \eref{eqn:langevin} and by $\CP_t$ the semigroup
defined on finite measures by
\begin{equ}
\bigl(\CP_t\mu\bigr)(A) = \expect \bigl(\mu \circ \phi_t^{-1}\bigr)(A)\;.
\end{equ}
Since $\d_t + \LL$ is hypoelliptic, $\CP_t$ maps every measure into a measure with a
smooth density with respect to the Lebesgue measure. It can therefore be restricted
to a positivity preserving contraction semigroup on $\L^1(\T^n \times
\R^{n},dx\,dy)$. The generator $\tilde \LL$ of $\CP_t$ is given by the formal
adjoint of $\LL$ defined in \eref{eqn:generator}.

We now define an operator $K$ on $\L^2(\T^n \times \R^{n},dx\,dy)$ by closing the
operator defined on $\CC_0^\infty$ by
\begin{equ}[e:defK]
K = -e^{{\delta\over 2} \|y\|^{2}}\tilde \LL e^{-{\delta\over 2} \|y\|^{2}}\;.
\end{equ}
The operator $K$ is then given by
\begin{equs}
K &= -{\sigma^2 \over 2\tau} \Delta_y + {\delta \over \tau}\Bigl(1-{\delta \sigma^2
\over 2}\Bigr)\|y\|^2
+ {1\over \tau}\bigl(\delta\sigma^2-1\bigr)\Bigl(y\cdot\nabla_y + {n\over 2}\Bigr) \\
&\quad +{1\over \sqrt \tau} \bigl(y\cdot\nabla_x + v(x)\cdot\nabla_y\bigr) - {n
\over 2\tau}\;.
\end{equs}
Note at this point that $\delta < 2\sigma^{-2}$ is required to make the coefficient of
$\|y\|^2$ in this expression strictly positive. This can be written in H\"ormander's
``sum of squares'' form as
\begin{equ}
K = \sum_{i=1}^{2n} X_i^* X_i^{} + X_0\;,
\end{equ}
with
\begin{equs}[2]
X_i &= {\sigma\over \sqrt{2\tau}}\d_{y_i} & &\kern-10em\text{if $i=1\ldots n$,} \\[0.4em]
X_i &= \sqrt{{\delta \over \tau}\Bigl(1-{\delta \sigma^2 \over 2}\Bigr)}y_{i-n}&
&\kern-10em\text{if
$i=(n+1)\ldots 2n$,} \\[0.4em]
X_0 &= {1\over \tau}\bigl(\delta\sigma^2-1\bigr)\Bigl(y\cdot\nabla_y + {n\over
2}\Bigr)+{1\over \sqrt \tau} \bigl(y\cdot\nabla_x + v(x)\cdot\nabla_y\bigr) - {n
\over 2\tau}\;.
\end{equs}
Since $v$ is $\CC^\infty$ on the torus, it can be checked in a very straightforward
way that the assumptions of \cite[Thm.~5.5]{EH00} are satisfied with $\Lambda^2 = 1
- \Delta_x - \Delta_y + \|y\|^2$. Combining this with \cite[Lem.~5.6]{EH00}, we see
that there exists $\alpha > 0$ such that, for every $\gamma > 0$,
 there exista a positive constant $C$ such that
\begin{equ}[e:boundK]
\|\Lambda^{\alpha+\gamma} f\| \le C \bigl(\|\Lambda^{\gamma}Kf\| +
\|\Lambda^{\gamma}f\|\bigr)\;,
\end{equ}
holds for every $f$ in the Schwartz space. Looking at \eref{e:boundK} with $\gamma =
0$, we see that $K$ has compact resolvent. Since $e^{-{\delta \over 2}\|y\|^2}$ is an
eigenfunction with eigenvalue $0$ for $K^*$, it follows that $K$ has also an
eigenfunction with eigenvalue $0$, let us call it $g$. It follows from
\eref{e:boundK} and a simple approximation argument that $\|\Lambda^\gamma g\| <
\infty$ for every $\gamma > 0$, and therefore $g$ belongs to the Schwartz space.
Furthermore, an argument given for example in \cite[Prop~3.6]{EPR} shows that $g$ must be
positive. Since one has furthermore
\begin{equ}
\tilde \LL e^{-{\delta \over 2}\|y\|^2} g = 0\;,
\end{equ}
the function $\rho$ given by \eref{e:defrho} is the density of the invariant measure
of \eref{eqn:langevin}. This concludes the proof of Theorem~\ref{theo:boundIM}.
\end{proof}

Before we give bounds on \eref{eqn:cell}, we show the following little lemma.

\begin{lemma}\label{lem:kernel}
Let $\delta \in (0,2\sigma^{-2})$ and let $K$ be as in \eref{e:defK}. Then, the kernel
of $K$ is one-dimensional.
\end{lemma}

\begin{proof}
Let $\tilde g \in \ker K$. Then, by the same arguments as above, $e^{-{\delta \over
2}\|y\|^2} \tilde g$ is the density of an invariant signed measure for $\CP_t$. The
ergodicity of $\CP_t$ immediately implies $\tilde g \propto g$.
\end{proof}
Now we are ready to prove estimates on the solution of the Poisson equation
\eqref{e:pois}.
\begin{theorem}
\label{thm:boundphi}
Let $h \in \CC^{\infty}(\T^n \times \R^n)$ with $D_{x,y}^{\alpha} h \in L^2 (\T^n \times \R^n ; e^{-
\epsilon \|y \|^2} dx dy)$ for every multiindex $\alpha$ and every $\epsilon >0$. Assume further that
$\int h( x, y)\,\mu(dx\,dy) = 0$, where
$\mu$ is the invariant measure for \eref{eqn:langevin}. Then, there exists a function $f$ such that
\eref{e:pois} holds. Moreover, for every $\delta > 0$, the function $f$ satisfies
\begin{equ}[e:boundPhi]
f(x,y) = e^{{\delta\over 2} \|y\|^2} \tilde{f}(x,y)\;,\qquad \tilde{f} \in \cS\;.
\end{equ}
Furthermore, for every $\delta \in (0,2\sigma^{-2})$, $f$ is unique (up to an
additive constant) in  $L^2 (\T^n \times \R^n , e^{-
\delta \|y \|^2} dx dy)$.
 %the class of functions satisfying $\int |f(x,y)|^2 e^{-\delta
%\|y\|^2}\,dx\,dy < \infty$.
\end{theorem}

\begin{proof}
By hypoellipticity, if there exists a distribution $f$ such that \eref{eqn:cell}
holds, then $f$ is actually a $\CC^\infty$ function.

We start with the proof of existence. Fix $\delta \in (0,2\sigma^{-2})$, consider the
operator $K^*$ which is the adjoint of the operator $K$ defined in \eref{e:defK},
and define the function
\begin{equ}
u(x,y) = h(x,y)\, e^{-{\delta \over 2}\|y\|^2}\;.
\end{equ}
It is clear that if there exists $\tilde{f}$ such that $K^*\tilde{f} = u$, then
$f = e^{{\delta \over 2}\|y\|^2}\tilde{f}$ is a solution to \eref{e:pois}.
Consider the operator $K^* K$. By the considerations in the proof of
Theorem~\ref{theo:boundIM}, $K^* K$ has compact resolvent. Furthermore, the kernel
of $K^*K$ is equal to the kernel of $K$, which in turn by \lem{lem:kernel} is equal
to the span of $g$. Define $\CH = \scal{g}^\bot$ and define $M$ to be the
restriction of $K^* K$ to $\CH$. Since $K^*K$ has compact resolvent, it has a
spectral gap and so $M$ is invertible. Furthermore, since $\LL y = \tau^{-1/2} v(x)
- \tau^{-1} y$, one checks easily that $f \in \CH$, therefore $\tilde{f} = K
M^{-1} u$ solves $K^*\tilde{f} = u$ and thus leads to a solution to
\eref{e:pois}.

Since $K^*$ satisfies a similar bound to \eref{e:boundK} and since $\|\Lambda^\gamma
u\|<\infty$ for every $\gamma > 0$, the bound \eref{e:boundPhi} follows as in
Theorem~\ref{theo:boundIM}. The uniqueness of $u$ in the class of functions under
consideration follows immediately from \lem{lem:kernel}.
\end{proof}
\begin{remark}
Note that the solution $f$ of \eqref{e:pois} is probably not unique if we allow for functions that
grow faster than $e^{\sigma^{-2}\|y\|^2}$.
\end{remark}
% We can, however, prove, uniqueness of solutions to the
%cell problem in the appropriate Hilbert space. Let $L^2_{\mu}$ be the set of functions in
%$L^2(\mu)$ which are centered with respect to the invariant measure. We will denote the inner
%product in this space by $\langle \cdot,\cdot \rangle_{0}$. Let now $\HH_1$ denote the set of
%functions in $L^2_{\mu}$ whose gradients with respect to $y$ also belong to $L^2_{\mu}$. We equip
%this space with the norm
%
%\begin{eqnarray}
%\|f \|^2_1 & = & \langle f, (I -\LL) f \rangle_{\mu}
%
%           \nonumber \\ & = &
%
%                  \|f \|_0^2 + \frac{\sigma^2}{2 \tau}\| \nabla_y f \|_0^2\;.
%
%\label{e:h1_nm}
%
%\end{eqnarray}
%
%Now we can prove uniqueness of solutions in the space $\HH_1$.
%
%\begin{lemma}
%
%\label{lem:uniq}
%
%There exists a unique solution of \eqref{e:pois} in $\HH_1$.
%
%\end{lemma}
%
%\proof Assume that there are two solutions $\Phi_1$ and $\Phi_2$ and denote their
%difference by $r$. Then $r$ satisfies the homogeneous equation
%
%$$
%-\LL r = 0
%$$
%
%from which it follows that $\|\nabla_y r \|_0 = 0$ and hence that $r$ is independent
%of $y$. Thus, equation $\LL r = $ becomes
%
%$$
%y \cdot \nabla_x r = 0.
%$$
%
%This equation should hold for every $y \in \R^d$ from which it follows, since $r$ is
%mean zero, that $r = 0$. \qed
%
\begin{remark}
The identity $y \tilde{\LL} \rho = 0$, where $\tilde{\LL}$ is the formal adjoint of
$\LL$, immediately yields that $\int y \,\mu(dx, dy) = \sqrt{\tau} \int v(x) \,\mu(dx, dy)$.
In particular, the assumption that the drift is centered implies that $y$ is also
centered. Moreover, $y$ clearly satisifies the smoothness and fast decay assumptions of Theorem
\ref{thm:boundphi}. Hence, the theorem applies to each component of
equation \eqref{eqn:cell} and we can conclude that there exists a unique smooth vector valued
function $\Phi$ which solves the cell problem and whose components satisfiy estimate
\eqref{e:boundPhi}.
\end{remark}
\subsection{Estimates on the particle velocity}
\label{sec:vel}
One has the following bound
\begin{lemma}\label{lem:expbounds}
There exists a constant  $\gamma > 0$ such that
\begin{equs}
\expect \exp\Bigl({1\over 2}\|\sigma^{-1}y(t)\|^2\Bigr) &\le \exp \Bigl({1\over 2}\|\sigma^{-1}y(0)\|^2 + \gamma t\Bigr)\;,\\
\expect \exp\Bigl({1\over 8\tau}\int_0^t \|\sigma^{-1}y(s)\|^2\,ds\Bigr) &\le \exp
\Bigl({1\over 4}\|\sigma^{-1}y(0)\|^2 + {\gamma\over 2} t\Bigr)\;.
\end{equs}
holds for any initial condition $y(0)$ and every $t > 0$.
\end{lemma}

\begin{proof}
It\^os formula yields immediately the existence of a constant $\gamma$ such that
\begin{equs}
{1\over 2}\|\sigma^{-1}y(t)\|^2 &\le {1\over 2}\|\sigma^{-1}y(0)\|^2 + \gamma t \\
&\quad - {1\over 2\tau}\int_0^t \|\sigma^{-1}y(s)\|^2\,ds + {1\over \sqrt\tau }
\int_0^t \scal{\sigma^{-1}y(s),d\beta(s)}\;.
\end{equs}
The first bound follows by exponentiating both sides and taking expectations. The
second bound follows in a similar way after dividing both sides by $2$.
\end{proof}

This yields the following:
\begin{theorem}
Let $\psi:\T^n\times \R^{n} \to \R$ be such that
\begin{equ}
\sup_{x \in \T^n,  y\in\R^n} \Bigl|\psi(x,y)\exp \Bigl(-{1\over
4}\|\sigma^{-1}y\|^2\Bigr)\Bigr| < \infty\;.
\end{equ}
Then, there exist constants $C, \delta > 0$ such that
\begin{equ}[e:boundconv]
\expect \bigl(\psi(x(t),y(t))\bigr)- \int_{\T^n \times \R^n}\psi(x,y)\,\mu(dx,dy)
\le C \exp \bigl({\|\sigma^{-1}y(0)\|^2-\delta t}\bigr)\;.
\end{equ}
\end{theorem}

\begin{proof}
From the smoothing properties of the transition semigroup associated to
\eref{e:system}, combined with its controllability and the fact that $\|y\|^2$ is a
Lyapunov function, one gets the existence of constants $C$ and $\delta'$ such that
\begin{equ}
\|\CP_t(x,y;\cdot\,) - \mu\|_\TV \le C (1 + \|y\|^2 ) e^{-\delta' t}\;.
\end{equ}
(See \textit{e.g.}\ \cite{MT} for further details.). Here $\|\nu - \mu\|_\TV$ denotes the total
variation distance between the measures $\mu$ and $\nu$. Cauchy-Schwarz furthermore
yields the generic inequality
\begin{equ}[e:boundgen]
\Bigl|\int f\,d\mu - \int f\,d\nu\Bigr| \le \sqrt{\|\mu-\nu\|_\TV \int f^2\,(d\mu +
d\nu)}
\end{equ}
The bound \eref{e:boundconv} immediately follows by combining \lem{lem:expbounds}
with \eref{e:boundgen}.
\end{proof}

We also have a much stronger bound on the supremum in time of the solution:
\begin{lemma}\label{lem:expbound}
For every $\kappa > 0$ and every $T > 0$, there exist constants $\delta, C > 0$ such
that
\begin{equ}
\expect \sup_{t \in [0,T\eps^{-2}]} \exp \bigl(\delta \|y(s)\|^2\bigr) \le
C\eps^{-\kappa}e^{\delta \|y(0)\|^2}\;,
\end{equ}
holds for every $\eps \in [0,1]$.
\end{lemma}

\begin{proof}
Let $\tilde y$ be the Ornstein-Uhlenbeck process defined by
\begin{equ}
\tilde y(t) = {1 \over \sqrt\tau}\int_0^t e^{-{t-s\over \tau}}\sigma\,d\beta(s)\;.
\end{equ}
Then (see \eg\ \cite{adler}), there exists constants $c_1$ and $c_2$ such that
\begin{equ}
\prob \Bigl(\sup_{t \in [s,s+T]} \|\tilde y(t)\| > \lambda\Bigr) \le c_1 e^{-c_2
\lambda^2}\;,
\end{equ}
for every $s > 0$. This immediately yields
\begin{equ}
\prob \Bigl(\sup_{t \in [0,T\eps^{-2}]} \|\tilde y(t)\| > \lambda\Bigr) \le c_1
\eps^{-2} e^{-c_2 \lambda^2}\;,
\end{equ}
which in turn implies that there exist constants $c_3$ and $c_4$ such that
\begin{equ}
\expect \Bigl(\sup_{t \in [0,T\eps^{-2}]} \exp \bigl(c_3 \|\tilde
y(t)\|^2\bigr)\Bigr) \le c_4 \eps^{-2}\;.
\end{equ}
The claim follows immediately by choosing $\delta = (c_3 \kappa)/2$ and by noticing
that there exists a constant $c_4$ such that $\|y(s)\| \le \|\tilde y(s)\| +
\|y(0)\| + c_4$ for all $s > 0$ almost surely.
\end{proof}
\section{Proof of Theorem \ref{thm:conv}}
\label{sec:proof}
In this section we prove Theorem~\ref{thm:conv}.
\begin{proof}
By Theorem \ref{thm:boundphi} we have $\Phi(y,z) \in \CC^{\infty}(\T^n \times \R^n,\R^n)$,
so we can apply the It\^{o} formula to the function $\Phi \bigl(\y(t), \x(t) \bigr)$
to obtain:
\begin{equs}
  \Phi \bigl(\y(t), \x(t) \bigr) -
\Phi(y , x)
                          &=
\frac{1}{\epsilon^2} \int_0^t \LL \Phi \bigl(\y(s), \x(s) \bigr) \, ds
+
\frac{1}{\epsilon} \frac{\sigma}{\sqrt{\tau}} \int_0^t \nabla_y \Phi \bigl(\y(s),
\x(s) \bigr) \, d \beta^\eps(s)
 \\ & =
 - \frac{1}{\epsilon^2} \frac{1}{\sqrt{\tau}} \int_0^t \y(s)  \,
 ds
+
 \frac{1}{\epsilon} \frac{\sigma}{\sqrt{\tau}} \int_0^t \nabla_y \Phi
  \bigl(\y(s), \x(s) \bigr)  \, d \beta^\eps(s)\;,
\end{equs}
where we defined $\beta^\eps(t) = \eps \beta(\eps^{-2} t)$ and we used \eqref{eqn:cell} to get the second line. We also interpret $\nabla_y \Phi$ as a linear map from $\R^n$ into $\R^n$.
Thus we have:
\begin{equs}
 \xe(t) & =  \epsilon x + \frac{1}{\epsilon} \frac{1}{\sqrt{\tau}} \int_0^t \y(s)\, ds \\
&= \epsilon x   - \epsilon \bigl( \Phi \bigl(\y(t), \x(t) \bigr) - \Phi (y , x )
\bigr) +
 \frac{\sigma}{\sqrt{\tau}} \int_0^t \nabla_y \Phi\bigl(\y(s), \x(s) \bigr)  \, d \beta^\eps(s) \\
& =:
                    \epsilon x + \epsilon I^{\epsilon}_1(t) + M^{\epsilon}(t)\;.
\label{eqn:x_cell}
\end{equs}

It follows from \eref{e:boundPhi} and \lem{lem:expbound} that, for every $p > 0$ there
exists a constant $C$ such that
\begin{equ}
\expect \sup_{t \in [0,T]} \bigl|I^{\epsilon}_1(t)\bigr|^p \le C \eps^{-{p\over 2}}\;.
\end{equ}
It is therefore sufficient to show that \eref{eqn:est} and \eref{e:strong} hold with
$\mu_\eps$ replaced by the law of the martingale term $M^\eps$. We first show that
\eref{eqn:est} holds. This is equivalent to showing that, for every $\ell \in \BB^*$ one can
construct a random variable $B_\ell$ such that
\begin{equ}[e:requested]
\expect |B_\ell - \ell(M^\eps)|^p \le C \eps^{\alpha p}\;,
\end{equ}
holds uniformly over $\|\ell\|\le 1$, and such the law of $B_\ell$ is given by $\ell^*\mu$.
We therefore fix $\ell \in \BB^*$ with $\|\ell\|\le 1$, which we interpret as a $\R^n$-valued
measure with total mass (\textit{i.e.}\ the sum of the masses of each of its components) smaller than $1$.
We also use the notation $\ell_t = \ell([t,T])$.

Integrating by parts, we can write
\begin{equ}
\ell(M^\eps) = \int_0^T \scal{M^\eps(t), \ell(dt)} =  {\sigma \over \sqrt\tau}\int_0^T \scal{\ell(t), \nabla_y \Phi \bigl(\y(t), \x(t)\bigr)\,d\beta^\eps(t)}\;.
\end{equ}
We now define on the interval $[0,T]$ the $\R$-valued martingale $M_\ell^\eps$ by
\begin{equ}
M_\ell^\eps(t) = {\sigma \over \sqrt\tau}\int_0^t \scal{\ell(s), \nabla_y \Phi \bigl(\y(s), \x(s)\bigr)\,d\beta^\eps(s)}\;.
\end{equ}
According to the Dambis--Dubins--Schwartz theorem \cite[Thm~1.6]{yor} there exists a
Brownian motion $B$ such that $M_\ell^{\epsilon}(t)$ can be written as
\begin{equ}
M_\ell^{\epsilon}(t) = B \bigl(\langle M_\ell^{\epsilon}, M_\ell^{\epsilon} \rangle_t \bigr) = B
\Bigl({\sigma^2 \over \tau} \int_0^t \scal[B]{\ell(s), \bigl(\nabla_y \Phi \otimes  \nabla_y \Phi\bigr) \bigl(\y(s),\x(s)\bigr)\, \ell(s)}\, ds
\Bigr)\;.
\end{equ}

On the other hand, the measure $\ell^*\mu$ is a centered Gaussian measure with
variance $ \int_0^T \scal[b]{\ell(s), \KK^2 \ell(s)}\,ds$, so we can choose $B_\ell$
to be given by
\begin{equ}
B_\ell = B_\ell^T\;,\qquad B_\ell^t = B\Bigl( \int_0^t \scal[b]{\ell(s), \KK^2
\ell(s)}\, ds \Bigr)\;.
\end{equ}
We will actually show a stronger bound than \eref{e:requested}, namely we will show that
\begin{equ}[e:strongb]
J_\eps^p := \expect \sup_{t \in [0,T]} |B_\ell^t - M_\ell^{\epsilon}(t)|^p \le C \eps^{\alpha p}\;.
\end{equ}
We use the H\"{o}lder continuity of the Brownian
motion $B$, together with the Cauchy--Schwarz inequality to derive the estimate
\begin{equs}
J_\eps^p
& \leq  \expect \Bigl(\mbox{H\"{o}l}_{\alpha}^{p}(B) \sup_{0\le t\le T}\Bigl|
\int_0^t \scal[B]{\ell(s), \Bigl({\sigma^2 \over \tau} \bigl(\nabla_y \Phi \otimes  \nabla_y \Phi\bigr) \bigl(\y(s),\x(s)\bigr) - \KK^2\Bigr)\,\ell(s)}\,ds
\Bigr|^{\alpha p} \Bigr) \\
& \le
                \bigl(\expect\, \mbox{H\"{o}l}^{2p}_{\alpha}(B)\bigr)^{\frac{1}{2}}
                \Bigl( \expect\sup_{0\le t\le T}\Bigl|\int_0^t \scal[b]{\ell(s), H\bigl(\y(s),\x(s)\bigr)\,\ell(s)} \,ds \Bigr|^{2\alpha p} \Bigr)^{1\over 2}\\
& \le
                C
                \Bigl( \expect\sup_{0\le t\le T}\Bigl|\int_0^t \scal[b]{\ell(s), H\bigl(\y(s),\x(s)\bigr)\,\ell(s)} \,ds \Bigr|^{2\alpha p} \Bigr)^{1\over 2}\;, \label{e:bestb}
\end{equs}
where we introduced the $n\times n$-matrix valued function
\begin{equ}
H(x,y) = {\sigma^2 \over \tau} \bigl(\nabla_y \Phi \otimes \nabla_y \Phi\bigr)(y , x) - \KK^2\;.
\end{equ}
In deriving the above estimate, we
have used the fact that for $\alpha < {1\over 2}$, the $\alpha$-H\"{o}lder constant of a Brownian motion is uniformly
bounded on every bounded interval \cite[Thm~2.1]{yor}.

Note now that since $\ell(t)$ is of bounded variation, $\ell(t) \otimes \ell(t)$ is also of bounded variation,
so there exists a $n\times n$-matrix valued measure $\tilde \ell$ on $[0,T]$ such that
$\ell(t) \otimes \ell(t) = \tilde \ell([t,T])$. Therefore, we can integrate by parts in \eref{e:bestb} to obtain
\begin{equs}
J_\eps^p &\le C
                \Bigl( \expect\sup_{0\le t\le T}\Bigl|\mbox{Tr} \int_0^t \int_0^s  H\bigl(\y(r),\x(r)\bigr)\,dr\,\tilde \ell(ds)  \Bigr|^{2\alpha p} \Bigr)^{1\over 2} \\
                &\le
                C\Bigl( \expect\sup_{0\le t\le T} \Bigl\| \int_0^t H\bigl(\y(s),\x(s)\bigr)\,ds  \Bigr\|^{2\alpha p} \Bigr)^{1\over 2}
\end{equs}

Consider now the Poisson equation
\begin{equation}
-\LL F = H\;.
\label{eqn:F}
\end{equation}
By the definition of $\KK^2$, we have $\int H(x,y)\,\mu(dx,dy) = 0$ (for each component), and we
furthermore have $\exp(-\delta \|y\|^2)H \in \cS$ for every $\delta > 0$. Therefore,
using the same reasoning as in the proof of Theorem~\ref{thm:boundphi}, equation
\eqref{eqn:F} has a unique smooth solution satisfying
\begin{equ}[e:boundF]
F(x,y) = e^{{\delta\over 2} \|y\|^2} \tilde\F(x,y)\;,\qquad \tilde\F \in \cS\;
\end{equ}
for every $\delta >0$. We can apply It\^{o} formula to deduce as before that
\begin{equ}
\int_0^t H(\y(s),\x(s)) \, ds  = - \epsilon^2 \bigl( F(\y(t), \x(t)) - F
 (y,x)\bigr) +  \frac{\epsilon}{\sqrt{\tau}} \int_0^t \nabla_y F(\y(s),\x(s)) \,
\sigma d \beta(s)\;.
\end{equ}
Therefore:

\begin{equs}
|J_\eps^p|^2 &\le
\eps^{4\alpha p} \expect \sup_{t\in[0,T]} \|F(\y(t), \x(t))\|^{2\alpha p} + C\eps^{2\alpha p}
\expect  \sup_{t\in[0,T]} \Bigl\|\int_0^t \nabla_y F(\y(s),\x(s)) \, d\beta(s)\Bigr\|^{2\alpha p} \;.
\end{equs}
Combining \lem{lem:expbound} with \eref{e:boundF}, the first term can be bounded by
\begin{equ}
\eps^{4\alpha p} \expect \sup_{t\in[0,T]} \|F(\y(t), \x(t))\|^{2\alpha p} \le C \eps^{-2\alpha p}\;.
\end{equ}
In order to control the second term, we use the Burkholder--Davis--Gundy inequality
followed by H\"older's inequality, assuming that $p > \frac{1}{\alpha}$:
\begin{equs}
\expect  \sup_{t\in[0,T]} \Bigl\|\int_0^t \nabla_y F(\y(s),\x(s)) \, d\beta(s)\Bigr\|^{2\alpha p}
&\le C \expect \Bigl(\int_0^T \|\nabla_y F(\y(s),\x(s))\|^2\,ds \Bigr)^{\alpha p} \\
&\le CT^{\alpha p-1} \sup_{t \in [0,T]} \expect  \|\nabla_y
F(\y(t),\x(t))\|^{2\alpha p}\;.
\end{equs}
This is bounded independently of $\eps$ by \eref{e:boundF} and \lem{lem:expbounds},
and so $J_\eps^p \le C \eps^{\alpha p}$, for $p > \frac{1}{\alpha}$.
When $p < \frac{1}{\alpha}$, one can bound $J_\eps^p$ using the higher
order moments. This completes the proof of bound \eref{e:requested} and thus
of the first part of \theo{thm:conv}.

The proof of the second part of \theo{thm:conv} is obtained in a straightforward way
as a particular case of \eref{e:strongb} if one makes the choice $\ell = k \delta_T$.
\end{proof}
\section{Conclusions}
\label{sec:concl}
The problem of homogenization for periodic hypoelliptic diffusions was studied in
this paper. It was proved that the rescaled particle position converges to a
Brownian motion with a covariance matrix which can be computed in terms of the
solution of the Poisson equation \eqref{eqn:cell}. Further, an upper bound on the
convergence rate in a suitable norm  was obtained. Our analysis is purely
probabilistic and this enables us to obtain more detailed information than what one
could obtain from studying the problem at the level of the Kolmogorov equation.

A very interesting question is whether a homogenization theorem of the form
\ref{thm:conv} holds for random drifts $v(x,t)$ and, if yes, under what conditions
on $v(x,t)$. From a mathematical point of view, it would be interesting to know
whether it is possible to achieve convergence in the $p$-Wasserstein distance for $n
> 1$. We plan to come back to these issues in a future publication.
%
%{\bf Acknowledgments}
%
\bibliography{mybib}

\begin{thebibliography}{EPRB99}
\expandafter\ifx\csname url\endcsname\relax
  \def\url#1{\texttt{#1}}\fi
\expandafter\ifx\csname urlprefix\endcsname\relax\def\urlprefix{URL }\fi

\bibitem[Adl90]{adler}
\textsc{R.~J. Adler}.
\newblock \emph{An {I}ntroduction to {C}ontinuity, {E}xtrema, and {R}elated
  {T}opics for {G}eneral {G}aussian {P}rocesses}, vol.~12 of \emph{Institute of
  Mathematical Statistics Lecture Notes---Monograph Series}.
\newblock Institute of Mathematical Statistics, Hayward, CA, 1990.

\bibitem[Bat85]{bhatta}
\textsc{R.~Battacharya}.
\newblock A central limit theorem for diffusions with periodic coefficients.
\newblock \emph{The Annals of Probability} \textbf{13}, (1985), 385--396.

\bibitem[BBJR95]{batty}
\textsc{C.~J.~K. Batty}, \textsc{O.~Bratteli}, \textsc{P.~E.~T. J{\o}rgensen},
  and \textsc{D.~W. Robinson}.
\newblock Asymptotics of periodic subelliptic operators.
\newblock \emph{J. Geom. Anal.} \textbf{5}, no.~4, (1995), 427--443.

\bibitem[BLP78]{lions}
\textsc{A.~Bensoussan}, \textsc{J.~Lions}, and \textsc{G.~Papanicolaou}.
\newblock \emph{Asymptotic analysis of periodic structures}.
\newblock North-Holland, Amsterdam, 1978.

\bibitem[CX97]{carmona}
\textsc{R.~Carmona} and \textsc{L.~Xu}.
\newblock Homogenization theory for time-dependent two-dimensional
  incompressible gaussian flows.
\newblock \emph{The Annals of Applied Probability} \textbf{7}, no.~1, (1997),
  265--279.

\bibitem[EH00]{EH00}
\textsc{J.-P. Eckmann} and \textsc{M.~Hairer}.
\newblock Non-equilibrium statistical mechanics of strongly anharmonic chains
  of oscillators.
\newblock \emph{Comm. Math. Phys.} \textbf{212}, no.~1, (2000), 105--164.

\bibitem[EPRB99]{EPR}
\textsc{J.-P. Eckmann}, \textsc{C.-A. Pillet}, and \textsc{L.~Rey-Bellet}.
\newblock Non-equilibrium statistical mechanics of anharmonic chains coupled to
  two heat baths at different temperatures.
\newblock \emph{Comm. Math. Phys.} \textbf{201}, no.~3, (1999), 657--697.

\bibitem[Fre01]{freidlin4}
\textsc{M.~I. Freidlin}.
\newblock On stable oscillations and equilibriums induced by small noise.
\newblock \emph{J. Statist. Phys.} \textbf{103}, no. 1-2, (2001), 283--300.

\bibitem[FW98]{freidlin1}
\textsc{M.~Freidlin} and \textsc{M.~Weber}.
\newblock Random perturbations of nonlinear oscillators.
\newblock \emph{Ann. Probab.} \textbf{26}, no.~3, (1998), 925--967.

\bibitem[FW99]{freidlin7}
\textsc{M.~Freidlin} and \textsc{M.~Weber}.
\newblock A remark on random perturbations of the nonlinear pendulum.
\newblock \emph{Ann. Appl. Probab.} \textbf{9}, no.~3, (1999), 611--628.

\bibitem[FW01]{freidlin5}
\textsc{M.~Freidlin} and \textsc{M.~Weber}.
\newblock On random perturbations of {H}amiltonian systems with many degrees of
  freedom.
\newblock \emph{Stochastic Process. Appl.} \textbf{94}, no.~2, (2001),
  199--239.

\bibitem[KO01]{komor_olla}
\textsc{T.~Komorowski} and \textsc{S.~Olla}.
\newblock On homogenization of time-dependent random flows.
\newblock \emph{Probab. Theory Related Fields} \textbf{121}, no.~1, (2001),
  98--116.

\bibitem[KV86]{kipnis}
\textsc{C.~Kipnis} and \textsc{S.~R.~S. Varadhan}.
\newblock Central limit theorem for additive functionals of reversible {M}arkov
  processes and applications to simple exclusions.
\newblock \emph{Comm. Math. Phys.} \textbf{104}, no.~1, (1986), 1--19.

\bibitem[LOY98]{landim_olla}
\textsc{C.~Landim}, \textsc{S.~Olla}, and \textsc{H.~T. Yau}.
\newblock Convection-diffusion equation with space-time ergodic random flow.
\newblock \emph{Probab. Theory Related Fields} \textbf{112}, no.~2, (1998),
  203--220.

\bibitem[MT93]{MT}
\textsc{S.~P. Meyn} and \textsc{R.~L. Tweedie}.
\newblock \emph{Markov chains and stochastic stability}.
\newblock Communications and Control Engineering Series. Springer-Verlag London
  Ltd., London, 1993.

\bibitem[Nel67]{nelson}
\textsc{E.~Nelson}.
\newblock \emph{Dynamical theories of {B}rownian motion}.
\newblock Princeton University Press, Princeton, N.J., 1967.

\bibitem[Nor94]{norris1}
\textsc{J.~R. Norris}.
\newblock Heat kernel bounds and homogenization of elliptic operators.
\newblock \emph{Bull. London Math. Soc.} \textbf{26}, no.~1, (1994), 75--87.

\bibitem[Nor97]{norris2}
\textsc{J.~R. Norris}.
\newblock Long-time behaviour of heat flow: global estimates and exact
  asymptotics.
\newblock \emph{Arch. Rational Mech. Anal.} \textbf{140}, no.~2, (1997),
  161--195.

\bibitem[Par99]{pardoux}
\textsc{E.~Pardoux}.
\newblock Homogenization of linear and semilinear second order parabolic pdes
  with periodic coefficients: A probabilistic approach.
\newblock \emph{Journal of Functional Analysis} \textbf{167}, (1999), 498--520.

\bibitem[PS03]{per_hom_inert_part}
\textsc{G.~A. Pavliotis} and \textsc{A.~Stuart}.
\newblock Periodic homogenization for inertial particles.
\newblock \emph{Preprint} .

\bibitem[PV85]{papan_varadhan}
\textsc{G.~Papanicolaou} and \textsc{S.~R.~S. Varadhan}.
\newblock Ornstein-{U}hlenbeck process in a random potential.
\newblock \emph{Comm. Pure Appl. Math.} \textbf{38}, no.~6, (1985), 819--834.

\bibitem[Rod89]{rodenh}
\textsc{H.~Rodenhausen}.
\newblock Einstein's relation between diffusion constant and mobility for a
  diffusion model.
\newblock \emph{J. Statist. Phys.} \textbf{55}, no. 5-6, (1989), 1065--1088.

\bibitem[RY99]{yor}
\textsc{D.~Revuz} and \textsc{M.~Yor}.
\newblock \emph{Continuous martingales and {B}rownian motion}, vol. 293 of
  \emph{Grundlehren der Mathematischen Wissenschaften [Fundamental Principles
  of Mathematical Sciences]}.
\newblock Springer-Verlag, Berlin, third ed., 1999.

\end{thebibliography}
%\bibliography{../bibtex_files/mybib}
\bibliographystyle{Martin}
\end{document}